# LiveRec: Prototyping Probes by Framing Debug Protocols

Jean-Baptiste Döderlein[a] , Riemer van Rozen[b] , and Tijs van der Storm[b,c]

a   ENS Rennes, Bruz, France
b   Centrum Wiskunde & Informatica (CWI), Amsterdam, The Netherlands
c   University of Groningen, Groningen, The Netherlands

## Abstract

**Context**  In the first part of his 2012 presentation "Inventing on Principle" [31], Bret Victor gives a demo of a live code editor for Javascript which shows the dynamic history of values of variables in real time. This form of live programming has become known as "probes" [3, 15, 18]. Probes provide the programmer with permanent and continuous insight into the dynamic evolution of function or method variables, thus improving feedback and developer experience.

**Inquiry**  Although Victor shows a working prototype of live probes in the context of Javascript, he does not discuss strategies for implementing them. Later work [18] provides an implementation approach, but this requires a programming language to be implemented on top of the GraalVM runtime [32]. In this paper we present LiveRec, a generic approach for implementing probes which can be applied in the context of many programming languages, without requiring the modification of compilers or run-time systems.

**Approach**  LiveRec is based on reusing existing debug protocols to implement probes. Methods or functions are compiled after every code change and executed inside the debugger. During execution the evolution of all local variables in the current stack frame are recorded and communicated back to the editor or IDE for display to the user.

**Knowledge**  It turns out that mainstream debug protocols are rich enough for implementing live probes. Step-wise execution, code hot swapping, and stack frame inspection provide the right granularity and sufficient information to realize live probes, without modifying compilers or language runtimes. Furthermore, it turns out that the recently proposed Debugger Adapter Protocol (DAP) [16] provides an even more generic approach of implementing live probes, but, in some cases, at the cost of a significant performance penalty.

**Grounding**  We have applied LiveRec to implement probes using stack recording natively for Java through the Java Debug Interface (JDI) [20], and through the DAP for Java, Python, C, and Javascript, all requiring just modest amounts of configuration code. We evaluate the run-time performance of all four probes prototypes, decomposed into: compile-after-change, hot swap, single step overhead, and stack recording overhead. Our initial results show that live probes on top of native debug APIs can be performant enough for interactive use. In the case of DAP, however, it highly depends on characteristics of the programming language implementation and its associated debugging infrastructure.

**Importance**  Live programming improves the programmer experience by providing immediate feedback about a program's execution and eliminating disruptive edit-compile-restart sequences. Probes are one way to shorten the programmer feedback loop at the level of functions and methods. Although probes are not new, and have been implemented in (prototype) systems, LiveRec's approach of building live probes on top of existing and generic debug protocols promises a path towards probes for a host of mainstream programming languages, with reasonable effort.

# The Art, Science, and Engineering of Programming

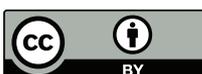

 **Introduction**

Live programming [7, 10, 15, 26] promises a better programming experience by bringing the execution of a program closer to the abstractions of the source code. By erasing the distinction between a program on the one hand, and its execution on the other, programmers may enjoy a more fluid experience, with immediate dynamic feedback after every code change.

Live programming has a rich history, both in academia and industrial research,[1] originating in the original Lisp systems (e.g., [24, 28]) and Smalltalk [9] which allowed dynamic replacement of code units (hot swapping) and inspection of program structures at run-time (sometimes referred to as "programming in the debugger"). However, live programming became center stage after Bret Victor's influential talk "Inventing on Principle" in 2012.

In that presentation Victor makes a compelling case that (paraphrasing) "creators need to see what they are creating". In the case of programming, this entails continuous inspection of the run-time behavior of a program, and immediate and seamless adaptation after a code change. While all the prototypes demonstrated by Victor are impressive, for the purpose of this paper we zoom in on one of them: a live editor for Javascript with a side pane showing the history of values of local function variables during an execution (see Figure 1 below). Although not mentioned by Victor himself, this live programming feature has become known as "probes" [15, 18].

The question left unanswered in Victor's presentation is: how to build such probes? While there has been research showcasing implementation approaches [18], they depend on a specific run-time system (e.g., GraalVM), or employ language specific implementation tricks [25], and hence cannot be easily transferred to other languages. So a more specific question is: how to engineer probes *generically*, without having to deeply modify a programming language's compiler and/or runtime?

In this paper we present LIVEREC, an approach to implement probes by reusing standard debugger protocols and APIs. Most language implementations support an API to facilitate the implementation of debugger UIs and other run-time tools (e.g., profilers, tracers, etc.). It turns out that such facilities are sufficiently powerful for implementing live probes. The key to the approach of LIVEREC is to use a scripted debugger to (re)execute a function or method of the debuggee, step by step, and at each step record a snapshot of the current stack frame. After the method has finished executing, the resulting stack recording is communicated back to the IDE and visualized alongside the code, matching stack recording entries to the source code using source location information.

We first detail how to implement probes using a native debugger interface, the Java Debug Interface (JDI). Then we generalize the approach to employ the recently proposed Debugger Adapter Protocol (DAP) [16], a generic interface layer to decouple IDE debug affordances and debugger servers for specific programming language implementations. This implementation, called DAPROBE, is showcased with probe

---

[1] For brief and incomplete overview of some of the history of live programming, see [30]

implementations in Java (again), C, Python, and Javascript. As a result, probes can be supported in any programming language that provides a reasonably complete implementation of the DAP, with only limited amount of configuration code. Our prototypes demonstrate that LiveRec allows us to obtain probes in languages that do not share any run-time infrastructure at all.

Since live programming (and hence probes) is all about immediate feedback, it is important that the performance overhead of LiveRec's strategy does not impede practical use. We therefore provide a preliminary analysis and measurement of the performance of (re)compilation, hot swapping, step-wise execution, and stack recording. To assess the overall overhead of LiveRec, we furthermore decompose a small editing scenario (derived from Victor's presentation) in 19 logical steps, replay it on every probe implementation, and measure the performance of each edit step.

This paper is further organized as follows. We analyze the requirements for implementing probes in Section 2, and discuss the scope and baseline assumptions of the paper in Section 3. Section 4 then provides an overview of LiveRec and introduces *stack recording* as a technique to obtain the required information through standard debug protocols. The implementation of LiveRec on top of the Java Debugger Interface (JDI) is described in Section 5. We then show how to generalize LiveRec in the context of the Debugger Adapter Protocol (DAP) in Section 6 by presenting DaProbe, a generic base component for DAP-based probes. We provide an initial assessment of LiveRec in Section 7, demonstrating feasibility of LiveRec by implementing probes for Java, Python, C and Javascript using DaProbe. We provide preliminary performance results of the individual steps, and reproduce an editing scenario to assess practicality. The paper is concluded with a discussion of limitations (Section 8) and related work (Section 9), and concluding remarks with directions for further research.

## 2 Background and Overview

Programmers can use probes to assess the behavior of a single function (or method) under investigation. By subjecting this function to a series of inputs, and observing associated outputs together with the values of local variables over time, a programmer can immediately see the impact minor code changes have on the behavior.

The basic concept of function probing is illustrated in Figure 1. The left pane shows an editor containing the function's source code, in this case an implementation of binary search in Javascript. The right pane shows the values the local variables have next to the source lines they were last updated. The top two rows specify example input data that the function is executed on. The consecutive values of variables that are updated in a loop (e.g. low, high, mid and value) are displayed in sequence hirozontally. So for low, the history of values is 0, 3, and 5. Apparently, the update of high in the else-if branch has never executed since there is no (redundant) value display in the right pane. Finally, the last line shows the return value of the function. We learn that the key was not found in the array.

```
function binarySearch (key, array) {

    var low = 0;
    var high = array.length - 1;

    while (low <= high) {

        var mid = floor((low + high)/2);
        var value = array[mid];

        if (value < key) {
            low = mid + 1;
        }
        else if (value > key) {
            high = mid - 1;
        }
        else {
            return mid;
        }
    }

    return -1;

}
```

```
key = 'g'
array = ['a','b','c','d','e','f']
  low = 0
 high = 5

  low = 0  |  3  |  5
 high = 5  |  5  |  5
  mid = 2  |  4  |  5
value = 'c' | 'e' | 'f'

  low = 3  |  5  |  6

return -1
```

■ **Figure 1** Screenshot of Bret Victor's 2012 talk "Inventing on Principle" (23rd minute) [31]

What makes this process *live* is that while the programmer is editing the program on the left (or the input arguments to the function), the value display on the right continuously updates to make it apparent what happens at run time.

Implementing programming environments with probe functionality is non-trivial. Dissecting the example leads to the following technical requirements:

- there should be a way to specify a function's input arguments (e.g., key and array);
- after every change to the code, the function needs to be recompiled and reexecuted;
- during execution, the values of local variables need to be recorded and linked to the source locations where they were updated;
- the resulting list of historical values should be displayed to the programmer.

To meet these requirements, we propose LiveRec, a language agnostic approach to probes, built on top of off-the-shelf debugging infrastructure. In particular, we introduce the Live Probe Server, to abstract the details of debug protocols behind a reusable and generic interface. The IDE communicates with the Live Probe Server whenever the code of a probed method is edited, and whenever its input values change.

The Live Probe Server's scripting harness simulates execution by issuing step-over commands until the probed function completes. To keep stack recording local to the current function, it never steps *into* a call to another function, but such calls are nevertheless still executed.

In between steps, the Live Probe Server takes snapshots of the current stack frame to record the values of local variables. These snapshots are stored in a data structure, called *stack recording*, which supports displaying the values in the IDE, for instance like the tabular style of Figure 1, inline [18, 25], or using another dedicated UI affordance.

To allow multiple functions to be probed in a single debug sessions, the debuggee is actually loaded into a "mock debuggee", the Keep Alive Agent (KAA). This driver program prevents unnecessary restarts of the debugger, and gives the Live Probe Server full control about its execution. The KAA itself simply consists of a main function with while-true-loop, halted on a breakpoint, and will be the host environment for executing functions.

## 3 Scope and Assumptions

The design and engineering space of probes is both deep and wide. There are a lot of trade-offs at play, ranging from the most bespoke implementation strategies to completely language parametric approaches. This section aims to clarify the place this paper takes by discussing its scope and baseline assumptions. For potential solutions or mitigations we refer to Section 8.

This paper considers probes from a live programming angle for mainstream, imperative and/or object-oriented languages, with readily available and mature implementations, either in the form of compilers, virtual machines, interpreters, or a combination of any subset of those three. Whereas probes would be valuable for languages that are less widely used (e.g., Clojure, Haskell, Prolog, to name a few), we focus on mainstream imperative languages, because precisely their similarities allow us to abstract over minor differences, and obtain a generic, language agnostic approach. We also do not specifically focus on dynamic programming languages (even though we include Javascript), in which live programming facilities have been much more common (e.g., Lisp, Smalltalk), and may in fact be much easier to implement. The challenge this paper aims to address is to engineer probes in language ecosystems where it would be, in fact, very hard to do this natively, yet which offer a variety of tools which could be "abused" for the task.

One such set of tools are debuggers. We assume that those language implementations are accompanied with source-level debuggers, which allow programmers to debug their code through debug services in the IDE. The standard arsenal of debugger services that we assume to be available are:

- Launch: run a program (the debuggee) in "debug mode".
- Step-over: execute a single statement, abstracting over intermediate method calls. Intermediate method calls *are* executed, but do contribute to granularity of stepping through code.
- Set breakpoint: instruct the debugger to halt the execution upon hitting a particular statement.
- Hotswap: reload modified code modules, such as functions, classes, files, etc., into the debuggee, without restarting.
- Evaluate expression: request the debugger to evaluate a sourcecode snippet in a particular execution context (e.g., object, stack frame, etc.) of the debuggee.

One feature we do not consider to be "standard" is watchpoints (also known as data breakpoints, in the context of DAP). A watchpoint is like a breakpoint, but instead of halting execution on a statement, execution is stopped when the value of a variable is modified. This could indeed provide a generic implementation strategy for probes: put watchpoints on all local variables, run the method to completion, and in the meantime collect the values of the variables. In certain sense, even, probes could be seen as "watchpoints with memory". Nevertheless, it turns out that watchpoints are far less commonly implemented than the debug services above. For instance, in the Java ecosystem, watchpoints are only supported on fields, and not on local variables. While the DAP offers interface definitions to implement watchpoints, they are only fully realized in the context of C (i.e., GDB). The Java DAP implementation has the same limitations as JDI (only fields); there are no Javascript or Python DAP implementations supporting watchpoints.

No code is an island. Although this paper studies probes in the context of small and isolated methods, we are aware that realistic code always executes in the context of other source code, method dependencies, libraries, etc. All these factors may influence the performance and scalability of our approach. Nevertheless, we would like to distinguish two things: method size, computational complexity, and size of the unit of hot swapping (typically, the compilation unit: class, file, module).

The first dimension influences the performance of LiveRec in that the larger the method, the more steps (typically) need to be performed by the debugger. Computational complexity has an even higher impact: it would probably not be very useful to probe a method implementing an exponential algorithm. The third dimension, unit of hot swap, affects probe performance in that the larger the compilation unit, the more overhead potentially is generated by compiling and reloading the code. Note however, that hot swap unit size does *not* affect the performance of stepping through code. In this initial work we delegate such considerations to future work, acknowledging that this implies our results cannot yet be generalized to realistic, industrial software development settings.

No invocation is an island. To invoke a procedure or function requires actual values for the formal parameters. In many probe implementations these values are specified as examples so that execution can be continuous. Nevertheless, executing code in imperative and OO languages depends on much more than just the parameters. In OO languages, methods need a self object: this requires a constructor invocation, to initialize field, and this constructor, in turn has formal parameters. Often the parameters are non-primitive, without a literal source notations. In that case initializing parameters requires the instantiation of classes, which have constructors with parameters, ... etc.

The same holds for global variables, which need to be initialized to meaningful values. Moreover, it holds for any state invariant that needs to be satisfied in order to invoke a method or procedure. In this paper we consider such complications out of scope. It is generally infeasible to ensure all required invariants, so we currently assume the dependencies of an invocation are limited to obtaining values for formal parameters. In the context of Java instance methods, we construct the self-object using simple heuristics.

Yet another complication that we presently gloss over is recursion. If a method is recursive, either directly or indirectly, there is the question what to do with arriving back in the probed method: should we limit the value display to the history of a single stack frame? Or should we include the values of all intermediate recursive stack frame histories as well? While our implementation could easily be changed to support either option, the fact that this also presents a problem of visualization, we currently deem this issue out of scope.

No application is an island. Every realistic software system interacts with the outside world through input and output. This can range to simple updates of the screen, to writing files on disk or other actions that can be considered dangerous. The continuous execution of methods is at odds with the fact that methods may not be reentrant because of such side-effects. In the present paper we do not provide any solution to this predicament, and delegate the responsibility to the programmer for not causing any serious harm.

No compilation is an island. For compiled languages, such as C or Java, and any other language that requires software to be build using build scripts and/or dependency management tools, we assume that a program is built accordingly. When the programmer runs the application in debug mode and makes use of probes, the code reloading after a source change is implemented in a custom fashion, through the debugger protocol of the language implementation. In other words we assume that a session with probes enabled is not affected by the (overhead of) the build system. Whether this is realistic, is out of scope of this paper and merits further research. It is however instructive to note that it is already possible to invoke hot swapping facilities from mainstream IDEs for many programming language implementations/debuggers, albeit with certain restrictions.

Live programming is about programmer experience (PX), a particular kind of user experience (UX). Therefore, probes require dedicated user interface affordances to support the programmer in a way that is not disrupting or confusing the programming activity. All our prototypes and demonstrations employ only the bare minimum of UI design, basically to demonstrate feasibility of the approach, without claiming any benefits in terms of UX. How the value histories should be conveyed to the user is of paramount importance for acceptance of probes in general, however outside the scope of this paper.

While this paper approaches live programming and probes from the use case of application programming, we note that other situations might benefit from probes as well. One area that could be investigated is the debugging of (unit) tests. In this case example data is already available, and probes could help to diagnose failing tests. However, the probe is then "one step removed" from the starting point (the test). Another area that could benefit from probes is basic programming education: since programming assignments are typically small and isolated, this could provide an ideal scale to apply probes.

## 4  Framing Debug Protocols

This section provides a high-level overview of the general architecture of LiveRec, and introduces the concept of stack recordings.

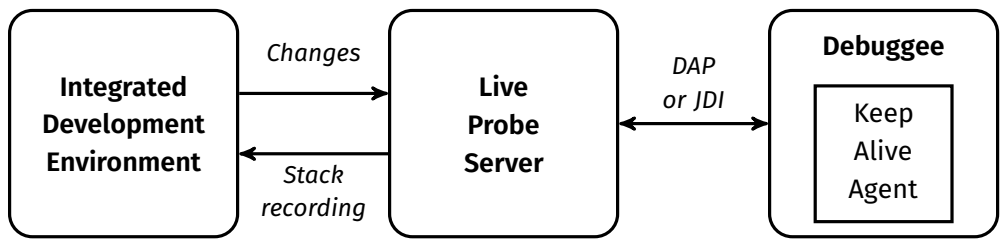

■ **Figure 2** General architecture

## 4.1 High-Level Architecture

A programming environment with probes must be able to react to two different events: a change in the code, or a change in the input data to the method. If the code changes, it needs to be reexecuted, possibly preceded by a compilation step in the case of compiled languages. Likewise, if one or more input values change, the probed method needs to be reexecuted, but in this case no recompilation is required. Both compilation (if any) of its enclosing compilation unit and reexecution of the probed method needs to be fast enough to avoid noticable delays in updating the display values of the method's variables.

The high-level architecture of LIVEREC is shown in Figure 2. The IDE is used by the programmer to edit source code. For any primitive edit, the programmer is editing at most one method, so there is at most one reload/re-execute operation to run. Of course, the cursor can be outside any method, but this is irrelevant to probes. If a method has been annotated with input values for its parameters (indicating that the programmer wants to probe the method), the code and the input values are sent to the Live Probes Server after every edit.

The server communicates with a debuggee through a custom scripting harness implemented using a debug protocol, such as JDI or DAP. This debugger component has been initialized with a dummy program, called the Keep Alive Agent (KAA). The code for a probed method is dynamically loaded into the KAA whenever possible, so that the method can be invoked through API calls of the debug protocol. By hot-swapping the code into the KAA we avoid a significant performance penalty of restarting the debugger itself every time a change is received by the Live Probes Server.

During execution of the probed method, the Live Probes Server collects all intermediate stack frame states in a stack recording, which is sent back to the IDE for display to the programmer.

## 4.2 Stack Recording

Displaying the dynamic evolution of variable values requires linking the consecutive values of each variable to their respective source locations. In LIVEREC this information is collected in *stack recordings*, which records the consecutive states of a single stack frame. In other words, a stack recording is a chained list of stack frame snapshots mapping the variables contained in them to the source location where they were modified. This data structure is illustrated in Figures 3 and 4.

```
1  //@foo(3)
2  int foo(int n) {
3    int i = 0;
4    while (i < n) {
5      i++;
6    }
7    return i;
8  }
```

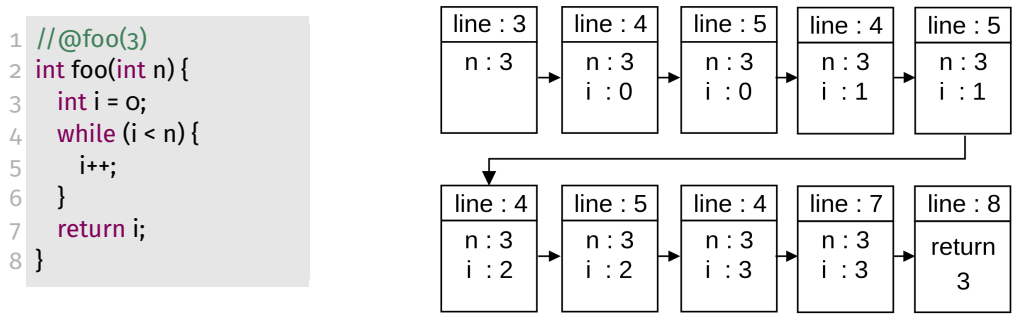

■ **Figure 3**  Stack Recording Example

| time → | $t_1$ | $t_2$ | $t_3$ | $t_4$ | $t_5$ | $t_6$ | $t_7$ | $t_8$ | $t_9$ |
|---|---|---|---|---|---|---|---|---|---|
| 1  //foo(3) | | | | | | | | | |
| 2  int foo(int n) { | | | | | | | | | |
| 3    int i = 0; | n=3 | | | | | | | | |
| 4    while(i < n) { | | n=3; i=0 | | n=3; i=1 | | n=3; i=2 | | | |
| 5      i++; | | | n=3; i=0; | | n=3; i=1 | | | | |
| 6    } | | | | | | | | | |
| 7    return i; | | | | | | | | n=3; n=3 | |
| 8  } | | | | | | | | | return 3 |

■ **Figure 4**  Tracking variable values across time and space (= source code)

Figure 3 displays a simple method annotated with input values using the special comment //@foo(3), indicating that the programmer wants to probe the method with input 3. The right side of the figure shows the corresponding stack recording. After every debug-step through the method, the current stack frame is cloned and added to the linked list, associating it with the source location (simplified as line numbers here) of the current statement.

Another way of visualizing stack recordings is shown in Figure 4, which details how stack recordings capture stepping through both time (= debug steps) and space (= source code). The right-hand part of the table can be "zipped" together, as it were, to obtain a tabular display like shown in Figure 1.

Stack recordings are simple and highly versatile: most mainstream programming languages feature a run-time stack and every debugger or debug interface has access to them.

Figure 5 illustrates how stack recordings are constructed using a UML message sequence diagram. After the probed method has been hot swapped into the KAA, the Live Probes Server sets a breakpoint on the first statement, and initiates its execution. It then continually steps through the method until the method completes. The stepping is performed using "step over" to avoid collecting stack frames of called methods. After each step the server asks for the current stack frame, and adds it to the stack recording, together with the current source location.

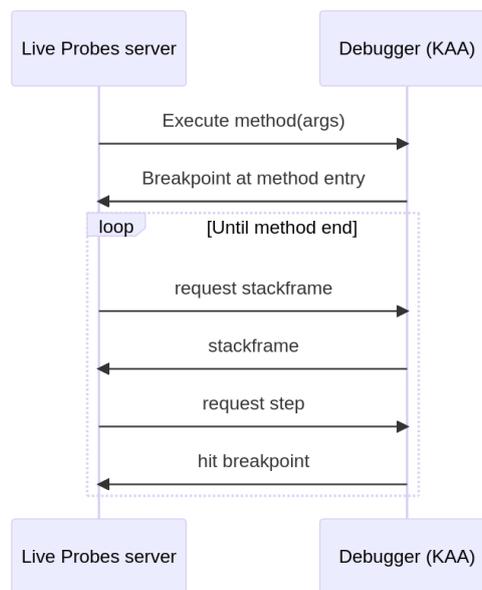

**Figure 5**   Record Stackrecording

## 5   Java Probes with JDI

**Listing 1**   Pseudo Java code capturing the gist of the Java Live Probes Server

```java
public class LiveProbesServer {
    KeepAliveAgent kaa = new KeepAliveAgent();

    public void run() {
        launchVM(kaa);
        setBreakpointOnWhileTrue();
    }

    public void loadClass(String className) { kaa.loadClass(className); }

    public StackRecording execute(Method method, Object[] arguments){
        setBreakpointAtMethod(method);
        invokeMethod(method, arguments);
        var stackRecording = new StackRecording();
        while (!isMethodFinished()) {
            stackRecording.add(getStackFrame(), getSourceLocation());
            stepDebuggee();
        }
        return stackRecording;
    }
}
```

■ **Listing 2** Simplified JDI Keep Alive Agent

```
1  public class KeepAliveAgent {
2    DynamicClassLoader dynamicClassLoader = ... // initialization omitted
3
4    public void loadClass(String className) { dynamicClassLoader.loadClass(className); }
5
6    public static void main(String[] args){ while (true) ; }
7  }
```

In this section we describe a Java implementation of the high-level approach introduced above, using the Java Debug Interface (JDI). Skeleton code of the key implementation classes are shown in Listing 1 (Live ProbesServer) and Listing 2 (Keep Alive Agent). The JDI is a complex API involving multi-threading and asynchronous computation (waiting for events etc.). The code snippets in the figures are therefore highly simplified, and function merely to give a flavor of how various components interact.

The LiveProbesServer (Listing 1) starts by initializing the KAA in a local field. Then, in the run method, a Java VM is launched with the KAA as the debuggee. When all the classes are loaded, *just* before main starts executing, a break point is set. (This is where the code snippet is a bit misleading; the real implementation involves waiting for an all-classes-loaded event, and then setting the breakpoint.) The breakpoint is needed to keep the debuggee in a state of suspension. This waiting state is necessary to use JDI reflective operations, such as newInstance (to instantiate classes), and invokeMethod (to call methods). Without the KAA the debug-enabled VM would need to be restarted after every code change, whereas now the changed code is hot swapped without starting a new debug session.

Whenever the programmer modifies a probed method in a class, the class is loaded into the KAA, using the loadClass method. Subsequently, the execute method is the trigger for probed method execution and stack recording, to be invoked by the IDE. Again, the code is simplified: we assume the probed method and its arguments have been parsed, compiled, and loaded into the VM, and hence are represented as reflective objects. A breakpoint is set at the start of the probed method (line 13), after which it is invoked. The while-loop then steps through the method as part of the KAA (which now contains the user code) until it completes, and then returns the collected stack recording.

Run-time class loading is implemented using a standard dynamic class loader, extended to support appending class paths at run time, in order to load the user code, which might reside in a location unknown to our scripting harness. Class loading is a lazy process: only if a dependency is needed it is loaded as well. If a dependency has been loaded previously, it is simply reused and not loaded again. The standard class loader infrastructure of Java does not support reloading classes if the class signature changes, which happens when fields or methods are added or removed, or when the inheritance hierarchy changes. Note however, that there are JVMs that do not suffer from those limitations (see, e.g., [12, 33]).

The above description of dynamic class loading for stack recording is sufficient for probing static methods, since all the required context for executing such a method is either static (as in static fields), or provided through method parameters. Ordinary methods require an instance of its enclosing class before they can be invoked. In our current implementation, the enclosing class is instantiated through the "smallest" constructor, measured by the number of constructor parameters. For each of the constructor parameters a default value is provided if the type is primitive, or `null` otherwise. The same strategy is used in the DAP Java implementation. We return to the issue of finding appropriate initializer arguments in Section 9.

## 6 DAPROBE: Implementing Probes through the DAP

The implementation of probes using JDI introduced in the previous section is language specific, because it is based on the native Java debug API JDI. The Debug Adapter Protocol (DAP) is a commonly accepted interface to communicate from an IDE to a programming language debugger. Designed to be used in modern IDEs like VS Code [17], the DAP offers a uniform interface for presenting a debugger user interface for any programming language implementing the DAP.

In this section we introduce DAPROBE, a generic probe server implemented in Python, which leverages the DAP for hot swapping, method execution, and stack recording. In Section 7 we evaluate this implementation for language implementations of C, Python, Java, and Javascript. This implementation is largely generic: all details regarding stack recording are handled language independently. A probe implementation for a specific language $X$ should provide implementations for the following extension points in the DAPROBE framework:

- A keep-alive agent in language $X$
- An implementation of how to reload code
- How to compile $X$ code (if applicable).
- Whether method execution is triggered by the debugger or in the debuggee itself.

Given language-specific implementations for the above four hooks, DAPROBE utilizes the DAP to provide probe functionality at modest implementation cost. The relevant DAP requests are summarized in Table 1.

We have instantiated DAPROBE for four languages: Java, C, Python, and Javascript. How the four extension points are realized is summarized in Table 2. Both the Java and C implementations compile code using their respective command line compilers.

The way the code is loaded also depends on the programming language. For Python, the code is loaded by being interpreted by the debug console during execution. For Javascript, a line is added at the end of the imported file to export all the functions in the file as a module, which is then loaded into the KeepAliveAgent with `require`. For Java, as with the JDI version, a Dynamic ClassLoader is used. For C, the code is loaded into a shared library that can be added and reloaded at run time.

Method execution can be divided into two categories. For DAP Java, Python and Javascript, calling methods directly from the debugger does not trigger breakpoints. To remedy this, execution must be initiated by the KeepAliveAgent rather than by a debugger command. To do this, the KeepAliveAgent code in these languages has

■ **Table 1** DAP requests used in implementing probes (see https://microsoft.github.io/debug-adapter-protocol/specification)

| Request | Usage |
| --- | --- |
| Initialize and Launch | Initialize the DAP server and launch the debuggee |
| SetBreakpoints | Set breakpoints at location in source code |
| StackTrace | Get the state of the current stacktrace |
| Scopes and Variables | Get the current scopes and variables in it |
| Evaluate | Evaluate expression in the debuggee |
| Next and Continue | Request to step or continue in the debuggee |

■ **Table 2** Language specific configuration of DAPROBE

| Implementation | Compile | Load Code | Execution Caller |
| --- | --- | --- | --- |
| DAP Java | javac | Dynamic ClassLoader | Debuggee |
| DAP Python | | From debug console | Debuggee |
| DAP C | gcc | Loaded as shared library | Debugger |
| DAP Javascript | | Imported as module | Debuggee |

fields for referencing a method and its arguments (see Appendix E for the example of Python); when this information is entered, the agent starts the execution. In C, as in Java with JDI, the method is called directly from the debugger.

## 7 Evaluation

In this section we evaluate LIVEREC, by demonstrating feasibility, assessing effort of implementation, and measuring the performance of probed method execution.

### 7.1 Demo: A Minimal Live Programming Environment

#### 7.1.1 Victor Style

We have developed a prototype dynamic programming environment to demonstrate probes for Java, C, Python, and Javascript. It is implemented as a simple web-based IDE, showing the source code editor on the left (based on CodeMirror 5 [11]), and the probe visualization on the right. Each time a change is made in the code editor, the code is sent to the server, which then tries to compile it. If the code contains a comment beginning with @, followed by a function call, the server attempts to create a stack recording of that function, with the parameters provided.

```
1 //@binary_search({1,2,3,4,5,6},6,9)                    1
2 int binary_search(int arr[], int length, int target) { 2 arr=int *[...]; length=6; target=9; -> return_value=-1
3     int left = 0;                                       3 left= 0
4     int right = length - 1;                             4 right= 5
5     while (left <= right) {                             5 left  = 0 | 3 | 5 | 6
6                                                         6 right = 5 | 5 | 5 | 5
7         int mid = (left + right) / 2;                   7 mid= 2 | 4 | 5
8         if (arr[mid] == target) {                       8
9             return mid;                                 9
10        } else if (arr[mid] < target) {                 10
11            left = mid + 1;                             11 left= 3 | 5 | 6
12        } else {                                        12
13            right = mid - 1;                            13
14        }                                               14
15    }                                                   15
16    return -1;                                          16
17 }                                                      17
```

■ **Figure 6**  Demo of the live programming environment for C.

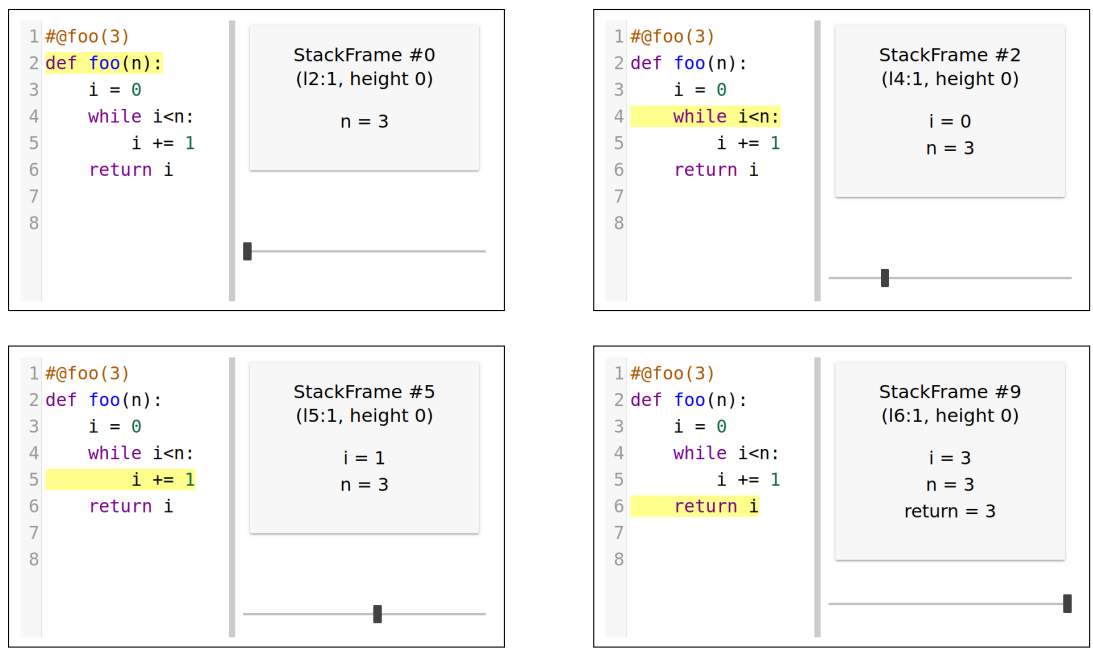

■ **Figure 7**  Exploring stack recordings: the slider moves through time and space of a single method execution.

Figure 6 shows a screen shot of a session using the C probe implementation. The left shows a binary search function, and the right panel displays the values of the probed variables.

### 7.1.2 Exploring Stack Recordings

We further have developed a stack recording exploration mode, allowing the programmer to travel in time through a method execution. Figure 7 shows four screen shots, each detailing a point in time in the history of the stack recording, namely, points 0, 2, 5, and 9. At each point in time the actual values of the variables are shown, and the corresponding point in space is highlighted in the editor in yellow. The slider at the bottom of each panel is used to move through time and space of the method.

■ **Table 3** Effort of implementing probes, measured in Source Lines of Code (SLOC)

| Implementation | #SLOC Probe Server | | #SLOC Keep-Alive Agent | |
|---|---|---|---|---|
| Java JDI | 521 | Java | 75 | Java |
| *DaProbe* | 270 | *Python* | — | |
| DAP Python | 142 | Python | 26 | Python |
| DAP Java | 360 | Python | 114 | Java |
| DAP Javascript | 256 | Python | 24 | Javascript |
| DAP C | 181 | Python | 14 | C |

For instance, the top left panel, shows the initial variable assignment just after the method has been invoked (the 0th snapshot of the stack frame); at this point in time, only n has a value, namely, 3. The top-right panel shows the stack frame after i = 0 has been executed, as visualized accordingly. The bottom right panel shows the final configuration, where the return value is shown as well.

### 7.2 Implementation Effort

Table 3 shows the number of lines of code for the server part and the KAA for each implementation, next to the implementation language. The DaProbe row corresponds to the base line code that is extended by the other DAP-based implementations. As can be seen from the table, probes can be implemented with very modest effort. Even the native JDI implementation requires less than 600 lines of code. The DaProbe framework itself requires a mere 270 lines of code. With the exception of DAP Java, the required configuration code to instantiate DaProbe is below 300 lines of code for all other languages. Appendices C, D, and E show the full implementations for the LiveRec base class (Python), the Python DAP extension, and the Python KAA, respectively.

### 7.3 Performance

In the context of live programming, it is essential to have short response times after user interactions. Hickups in the programmer's flow destroy the value that probes are supposed to provide in the first place. In this section we present preliminary performance results to appreciate the feasibility and scalability of LiveRec. Note however, that given our assumptions detailed in Section 3, no strong conclusions can as of yet be derived from the results. In particular, the presented figures paint a rather mixed picture, in which many aspects require further investigation and benchmarking.

We run micro-benchmarks to assess the overhead of using LiveRec to implement probes, to better understand the following questions:

- What is the performance cost of compiling and (re)loading code into the debugger (in s), depending on the size of the method (in LOC)?
- What is the performance of executing a method using consecutive debug-steps while constructing the stack recording (in s), depending on the number of debug-steps?

▪ What is the *overall* performance overhead (in s) during a realistic scenario of developing a method?

All experiments were carried out with OpenJDK 20.0.2 for Java, Python 3.11.2 for Python and Node 18.16.1 for Javascript. The C code was compiled using GCC 13.1.1 and GDB 13.2 was used for stack recording. The machine used to carry out the evaluations has an AMD Ryzen 5 2500U CPU and 8Gb of RAM.

### 7.3.1 Performance of Compiling and Loading Code

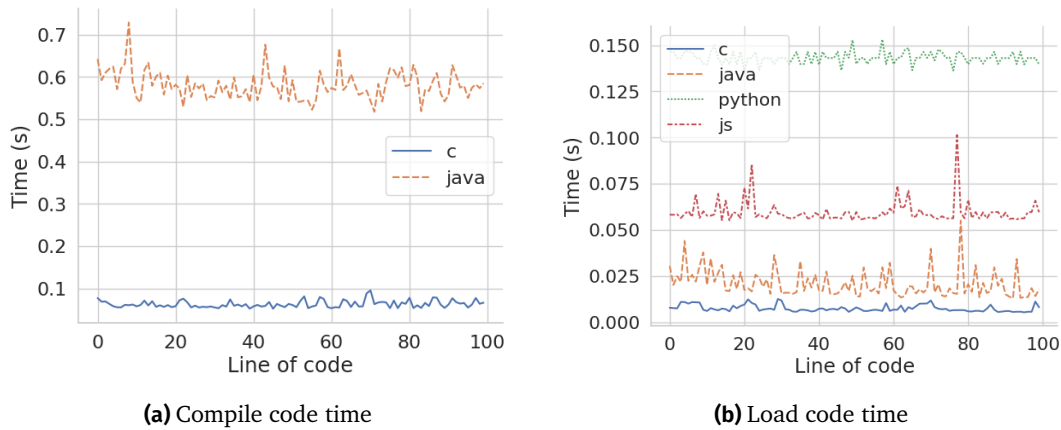

**(a)** Compile code time       **(b)** Load code time

■ **Figure 8** Performance of compiling and loading depending code size (with DAP)

In our evaluation, we assessed the time required to compile and load code into the debuggee for Python, C, Java and Javascript. To accomplish this, we compiled and loaded programs ranging from 5 to 100 lines of code. The summarized results can be found in Figure 8. Sub-figure 8a shows compilation time for Java and C relative to the number of lines of code. The compilation process was executed from the command-line using javac and gcc, respectively.

The compilation times seem to remain nearly constant, regardless of the number of lines of code, with an average of 34.6 ms for Java and 8.5 ms for C. Experiments on realistic code sizes are required however, to assess whether these numbers do not simply account for start-up costs of the respective compilers; naturally, in the limit, compilation time should depend on code size. Moreover, these measurements are based on compiling and loading an isolated function. While 100 lines of code seems a reasonable estimate for an average function or a method definition, we gloss over the fact such functions are typically contained in (much) larger modules or files (see also Section 3).

Sub-figure 8b depicts the loading time in the debugger for C, Java, Python and Javascript as a function of the number of lines of code. The data suggests that loading time remains constant irrespective of the number of lines of code. The same caveats mentioned above apply in this scenario as well: in the limit hot swapping should depend on the unit size of code that is loaded. One difference, however, is that hot swapping, by definition, does not require the operating system to start a new process.

### 7.3.2 Performance of Step-wise Execution

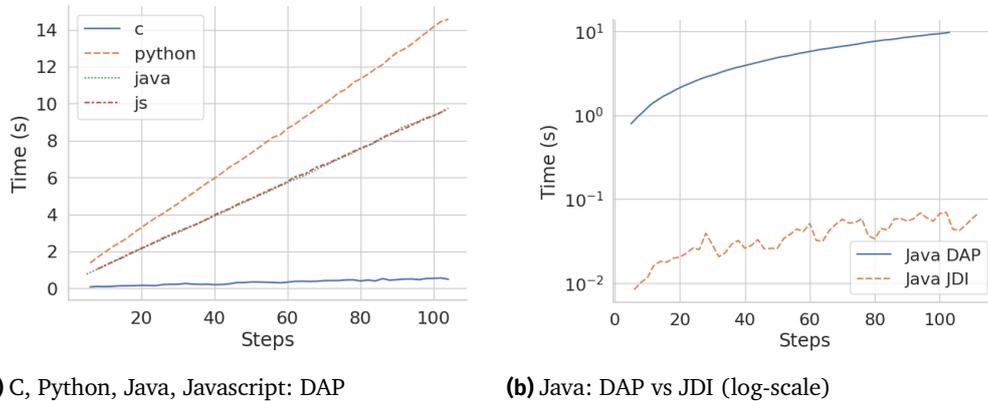

**(a)** C, Python, Java, Javascript: DAP

**(b)** Java: DAP vs JDI (log-scale)

■ **Figure 9**   Performance of stack recording per number of executed steps (in seconds)

Next, we look at the performance of probing as a function of the number of steps taken by the debugger. The results are shown in Figure 9, showing total execution time per number of steps during stack recording.

Figure 9a shows the results for the DAP implementations. The trend in this graph suggests a linear relationship with the number of steps taken by the probed function, which is to be expected, since obtaining a stack frame is a constant operation. The difference in time between the different DAP languages may be explained by the implementation of respective debuggers. For instance, in the case of C, the debug server is a simple wrapper around `gdb`, which has very good performance. Nevertheless, the results also seem to suggest that DAP itself adds considerable overhead: in an absolute sense the measured times are simply too high for viable probes.

The picture is different in the case of JDI, shown in Figure 9b, plotted against Java DAP measurements (in log-scale). There is a significant difference in performance between the two implementations, despite the fact that the DAP server implementation in Java also uses JDI internally. As a result, we hypothesize that indeed DAP overhead (JSON-RPC communication, cost of process context switching, serialization, etc.) is the culprit here. On a positive note, the JDI probes seem to at least stay well below the 1 second threshold.

### 7.3.3 Overall Performance of Probes

The measurements discussed above looked at the performance of isolated phases of going from a change in code to its execution using LIVEREC. Here we investigate the combined performance of all phases in a simple coding scenario. To do that we have recreated the live programming scenario presented in the video by Bret Victor, and executed it in all the probe implementations, measuring time after each code change.

The scenario is based on the development of a binary search function on an array of characters (cf. 1). It consists of 19 steps: 10 steps where the source code is modified and 9 steps where the input parameters are changed. The full scenario is included in Appendix A. Step 16 of the scenario causes the method to not terminate, so we set the maximum size of the stack recording at 80 snapshots, to avoid crashing the debugger. When the recording exceeds this limit, the debugger is stopped gracefully, and restarted. While 80 seems an arbitrarily low limit, it is easily made configurable through a configuration parameter. Since the code in Bret Victor's presentation is in Javascript, it has been ported to the other languages studied in this paper (see Appendix A).

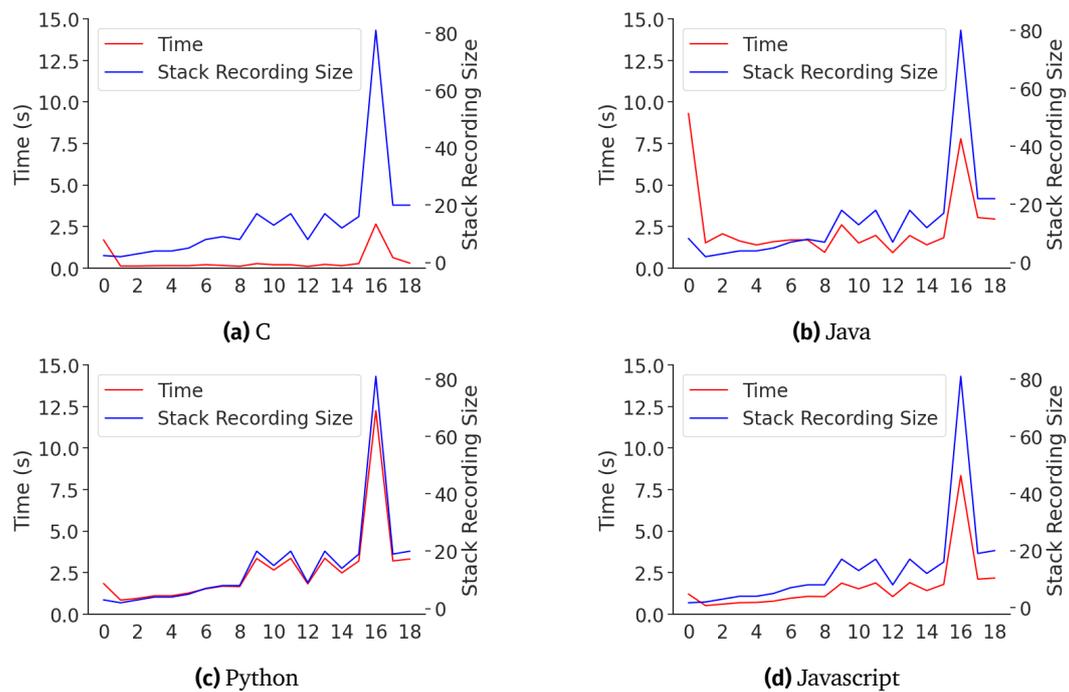

**Figure 10** Execution time and stack recording size for the binary search scenario

Figure 10 shows the time and number of stack frame snapshots recorded for each stage of the scenario in Python, C, Java and Javascript. When the initialisation is executed for Java and C, the time taken at each stage follows the same trend as the number of stack frame snapshots in the stack recording. At step 16, we also observe the case of an infinite loop in the scenario.

As can be seen from the plots, with the exception of C, the performance of probing method execution roughly floats between 1 and 3 seconds, which is above the 0.1 second limit that users of interactive systems begin to notice, so we have to conclude that the DAP based probe implementations incur too much performance overhead for realistic use.

### 7.4 Investigating DAP Overhead

The experiments above suggest that the DAP incurs a significant overhead. It is particularly visible comparing Java DAP to direct JDI usage, despite the fact that the Java DAP server also utilizes JDI internally. To understand the discrepancy better, we profiled the execution of the StackTrace request in the Java DAP implementation and measured the time spent in the DAPROBE client (implemented in Python) and in the Java DAP server separately.

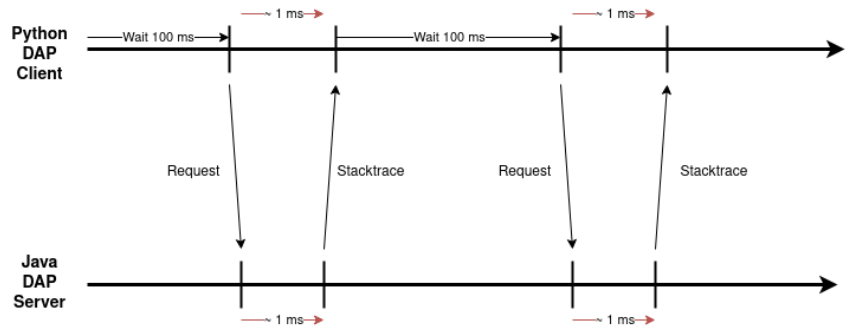

■ **Figure 11** Consecutive requests with pausing

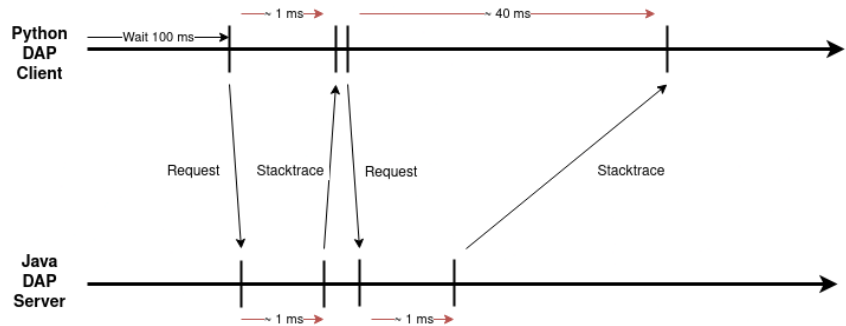

■ **Figure 12** Consecutive requests without pausing

The initial experiment stresses the Java DAP server with 200 requests interleaved with 100ms pauses, as illustrated in Figure 11. The measurements show that the time of one roundtrip is roughly 1ms. However, if the 100ms pauses are eliminated, even though it still requires 1ms in Java to send the response , the message arrives back in the client only after 40ms. This is shown in Figure 12.

These findings suggest that issue stems from the Java DAP server, or the DAP protocol layer, and not the DAPROBE client. We hypothesize two possible causes:

- Communication at the Java DAP level occurs via a socket, which might not flush quickly enough, causing delays.
- The Java DAP's integration into the Java Language Server Protocol (LSP) in a separate thread might lead to inefficient thread scheduling, contributing to the time lag.

Further investigations are needed to find the root cause of the performance penalty.

## 8 Discussion

While our approach to implement live probes on top of off-the-shelf debuggers promises to be a viable strategy to add live probes to many mainstream languages, some limitations and open questions remain. Below discuss these in more detail.

The abstraction boundaries provided by generic interfaces allow software developers to decouple clients and servers—yet that often creates other problems. This state of affairs can be recognized in some of the performance results of the DAP-based live probes discussed above. Layering another level of abstraction (the DAP) between the native debuggers and the live probe server, seems to cause performance to degrade, possibly due to serialization overhead, inter-process communication, and potential process context switches.

In our current approach, the programmer specifies input arguments to functions or methods using special comments in the code. This choice is motivated by the fact that it works in every programming language, does not require special UI affordances, and allows example data to be committed to version control systems. Nevertheless, while most primitive data types have a convenient, human readable literal notation, this can become cumbersome with complex, hetergeneous, structured data, such as arrays, objects, records, etc. Moreover, these literal notations are language specific: how to specify a constant array in C, Java or Javascript is very different, and needs to be parsed into actual data values from their textual representation in the special comments.

Another challenge is most visible in the case of object-oriented languages, such as Java, where methods typically live in classes, and can only be invoked when an instance of a class is available. This means the programmer has to specify not just the input parameters, but also values for fields, or arguments to constructors, so that the live probe server can construct the object. This situation becomes worse if, e.g., the constructor depends on other objects as argument, which, in turn, require initialization, and so on. Our current implementation constructs basic objects with default values for all constructor arguments. This could be extended to let the programmer gradually improve the binding of such parameters using more detailed and structured annotations.

The above problem of objects and initialization hints at a more basic and general question: how to obtain useful example data in the first place. In this paper we have assumed the programmer provides the data, but there are other strategies worth considering. For instance:

- Random value generation, as is employed in random testing [1, 5], could be used to assign values to parameters. However, in this case it is unclear if it leads to the insight the programmer wants, and would be very confusing if the values would change in every iteration.

- If a test suite is available, an off-line tool could observe frequently occurring values and objects passed into methods, through dynamic analysis of the test suite execution. Such common values could also be harvested from production runs. Just like with random generation, however, the problem is to make the assignments stable somehow.

- Finally, the search for values could be coverage guided, i.e. to use sets of values that maximize the path coverage of a function [13].

The above strategies absolve the programmer from having to specify the example values, but potentially at the cost of reduced quality of feedback. Further research is needed to find an appropriate middle ground between full manual specification, and fully automatic techniques.

Live probes require continual execution of methods and functions. This is fine if the code does not perform any harmful side-effects, such as sending out email, writing files, or launching missiles. While it is probably undecidable to statically check whether a piece of code might eventually perform such a side effect, we leave it to the programmer to indicate whether the live probing feature is switched on, e.g., by providing the annotation of the input arguments.

Possible future directions to solve this problem would include sandboxed execution, or intercepting all calls to IO libraries or the operating system. The latter technique would also allow the live probes to not only display variable values, but also, e.g., console output. A more disciplined solution would be offered by languages with built-in notions of capability-based security (such as, e.g., Newspeak [4]), so that the live probes can be configured to execute in an environment where dangerous features (such as IO) are not available.

While the probes that we have demonstrated support variables updated in loops by displaying them in tabular form, we have not discussed what happens with recursion. The probe execution repeatedly instructs the debugger to step-over function calls, so intermediate stack frames are automatically excluded from the stack-recording, *except* for when execution arrives back at the function that is currently being probed, since it has a breakpoint on entry. The current implementation simply continues, and adds stack frame snapshots to the recording, even though we are on a different level of the stack. Further research is needed to explicitly deal with this case, and find an appropriate tree-based or nested visualization.

## 9 Related Work

Live programming is a research area that studies how to make the programming activity more fluid and seamless, by eliminating mental gaps and improving immediate feedback [30]. The term refers to a wide array of techniques and approaches that crosscut Human Computer Interaction (HCI) and Programming Languages (PL) research. By providing immediate and continuous feedback, live programming environments improve code comprehension and potentially accelerate programming and debugging.

Tanimoto has introduced "levels of liveness" that indicate increasing amounts of feedback, responsiveness, interactivity, and predictive feed forward (design suggestions) [26]. LiveRec operates at level 3: informative, significant, and responsive. As an early example, Tanimoto describes VIVA a language for image processing [27]. VIVA expresses visual flows of image processing algorithms and communicates the effects on images in a human-friendly manner.

McDirmid was the first to study probes as part of a complete language design [15]. It is one of the live programming features that has seen the most adoption [22]. McDirmid describes a design to probes focused on debugging and problem diagnosis. Similar to LiveRec, his approach traces program executions that link stack frame states to source locations. The author demonstrates how this technique can be applied in a dedicated live programming environment for the YinYang language. In this case, the language *design* follows from the desire to have a better live programming experience. In the case of LiveRec, however, we assume the language design is fixed, and attempt to find a solution strategy that is a widely applicable to as many programming languages as possible.

Another example of probes was presented by Van der Storm and Hermans, as "Live Literals" [25]. In this case the resulting value histories of the probes are displayed inline in the source code itself *as actual source code*. The implementation is based on run-time instrumentation and origin tracking to update the literal source text in the editor.

Example-centric programming aims to add IDE support for examples to support the programming process [7]. Edwards demonstrates the "EG tool" an Eclipse plugin for example-centric programming in Java. Based on BeanShell, a custom JVM, EG tool provides live feedback about evaluated examples. LiveRec can be seen as a form of example-based programming, in the sense that the example inputs are guiding the probed method execution.

Niephaus et al. combine GraalVM and Truffle with the Language Server Protocol (LSP) to support example based programming for multiple languages [18]. Truffle is an interpreter framework designed for efficient language execution on the GraalVM Java virtual machine [32]. As far as we know, this is the only work offering a generic, reusable engineering approach for probes; nevertheless, it requires that the programming language runs on the GraalVM. LiveRec is more generic: we have demonstrated prototypes in languages that do not share any run-time infrastructure at all.

Omniscient debugging [2, 14, 19, 21], also known as back-in-time debugging, allows programmers travel back in time during debugging. Bousse et al. propose a reusable omniscient debugger for executable Domain-Specific Lanugages (DSLs) [2]. This is similar in aims to our work in that the omniscient debugging framework is language parametric. While not the explicit goal of LiveRec, the stack exploration tool demonstrated in Section 7 provides a limited, scoped form of time-travel, namely within the scope of a single method.

A much more ambitious approach to improving the construction of live programming systems with immediate feedback, is the Cascade meta-language [23]. The hypothesis of this work is that if one defines their language using Cascade, one obtains live run-time updates "for free". In particular, this approach solves the problem of migrating run-time state (e.g., the heap) after a program change [8, 29]. The key concept is that Cascade sees every change, both to the source code, and those triggered by users in the running application, as an edit transaction, thereby effectively fusing program change and user change. In the case of LiveRec, we assume the method execution is relatively stateless: if a method or function should depend on some form of ambient state (fields, global variables, etc.), then this state will be reset to its default values after hot swapping the code.

Executing methods in isolation requires input values. This topic has received attention in work on randomized testing (e.g., [1, 5, 6]) and symbolic and/or concolic execution (e.g., [13]). In both domains the challenge is twofold: input values are needed and functions have to be executed in isolation. Further research in these areas could help lift some of the assumptions that we made in Section 3.

## 10 Conclusion

Live programming technology promises to improve the programmer experience by providing immediate dynamic feedback about the execution of a program. Probes, as initially demonstrated by Bret Victor in his talk "Inventing on Principle", allow the programmer to observe the dynamic evolution of values of variables of a single method or function, for instance in the form of a tabular display, inline hints, or tool-tips. This display is *live*: after every code change or change in the example values, the display is updated immediately.

Probes have been researched before, but how to implement them for a wide range of languages has not received much attention. Either the implementations are bespoke, dependent on specific language designs or situations. Or, the implementation strategy depends on a specific run-time system, such as the GraalVM. In this paper we present a language-agnostic approach to probes. We show that by reusing facilities offered by debug protocols, we can implement live probes at low implementation cost.

We present LiveRec, an approach to record the stack frame evolution of a method's execution in a stack recording, which can be displayed in the user's IDE and linked to the statements of the code where the variable values were active. We demonstrate how this can be realized using the native Java Debug Interface (JDI), and the generic Debug Adapter Protocol (DAP). In the latter case we have implemented probes for C, Java, Python, and Javascript, with very modest language specific configuration code. An initial performance evaluation shows that native JDI probes are fast enough for interactive use; in the case of the DAP implementations, the performance is insufficient in most cases. Further research is needed to understand the precise reasons, and to find strategies to optimize.

LiveRec can be seen as a first attempt to adding probes to the standard arsenal of IDE features for mainstream languages. We hope it will provide a stepping stone to make probes as mainstream as C, Java, Python, and Javascript. Further research directions include: exploring probes for end-user programming environments and DSLs, further optimizing the DAP-based probes, and investigating how a first-class probe API can be integrated into the DAP or LSP.

**Acknowledgements**   We thank the anonymous reviewers for helpful comments in improving this paper, especially reviewer 3 was of great help.

## A  Binary Search Scenario

The full binary search programming scenario as derived from Bret Victor's presentation [31] (see also Figure 1; full source code snippets after stage 19 can be found in Appendix B):

1. The function is defined and the parameters are defined to be an array of characters( `['a','b','c','d','e','f']` ) and a target character( `'d'` ).

2. A new variable `low` is defined.

3. A new variable `high` is defined.

4. A new variable `mid` is defined.

5. The variable `mid` is changed to be a integer.

6. A new variable `value` is defined.

7. A if and else snippets is added.

8. The target is changed to `'b'` .

9. The target is changed to `'c'` .

10. The target is changed to `'d'` and the code from the `mid` is refactored to be in a `while(true)` loop.

11. The target is changed to `'a'` .

12. The target is changed to `'b'` .

13. The target is changed to `'c'` .

14. The target is changed to `'d'` .

15. The target is changed to `'e'` .

16. The target is changed to `'f'` .

17. The target is changed to `'g'` .

18. The condition of the while loop is changed to `low <= high` .

19. A return `-1` is added at the end of the function.

## B  Source Code of the Binary Search Function in Java, C, Python, and Javascript

### B.1  Java

```java
public class BinarySearch {
    public static int binarySearch(char[] array, char key) {
        int low = 0;
        int high = array.length - 1;

        while(low <= high){
            int mid = (low + high) / 2;
            char value = array[mid];

            if (value < key) {
                low = mid + 1;
            } else if (value > key) {
                high = mid - 1;
            } else {
                return mid;
            }
        }
        return -1;
    }
}
```

## B.2  C

```c
int binary_search(char arr[], int length, char target) {
    int left = 0;
    int right = length - 1;

    while(left <= right){
        int mid = (left + right) / 2;
        char value = arr[mid];

        if(value < target) {
            left = mid + 1;
        } else if(value > target) {
            right = mid - 1;
        } else {
            return mid;
        }
    }
    return -1;
}
```

## B.3  Python

```python
def binary_search(arr, target):
    left = 0
    right = len(arr) - 1

    while left <= right:
        mid = (left + right) // 2
        value = arr[mid]

        if value < target:
            left = mid + 1
        elif value > target:
            right = mid - 1
        else:
            return mid

    return -1
```

## B.4  Javascript

```javascript
function binary_search(arr, target){
    var low = 0;
    var high = arr.length - 1;
    while (low <= high) {
        var mid = Math.floor((low + high) / 2);
        var value = arr[mid];

```

```
 8            if (value < target) {
 9                low = mid + 1;
10            }
11            else if (value > target) {
12                high = mid - 1;
13            }
14            else {
15                return mid;
16            }
17        }
18        return -1;
19 }
```

### C  Abstract Base Class for Live Agents

```python
import subprocess
import os
from abc import ABC, abstractmethod
from typing import Any
from debugpy.common.messaging import JsonIOStream

from livefromdap.utils.StackRecording import StackRecording

class DebuggeeTerminatedError(Exception):
    def __init__(self):
        super().__init__("Debuggee terminated")

class BaseLiveAgentInterface(ABC):
    # omited for brevity

class BaseLiveAgent(BaseLiveAgentInterface):
    seq : int = 0
    debug : bool = False

    io : JsonIOStream

    def __init__(self, **kwargs : Any):
        self.debug = kwargs.get("debug", False)
        self.seq = 0

    def __del__(self):
        try:
            self.stop_server()
        except:
            pass

    def new_seq(self):
        self.seq += 1
        return self.seq

    def _handleRunInTerminal(self, output : dict):
        if output["type"] == "request" and output["command"] == "runInTerminal":
            # if not exists, create the tmp folder
            if not os.path.exists("tmp"):
                os.makedirs("tmp")

            debuggee = subprocess.Popen(
                output["arguments"]["args"],
```

```python
                    stdout=open("tmp/stdout.txt", "w"),
                    stderr=open("tmp/stderr.txt", "w")
                )
                process_id = debuggee.pid
                self.debugee = debuggee
                # send the response
                self.seq+=1
                response = {
                    "seq": int(output["seq"]) + 1,
                    "type": "response",
                    "request_seq": output["seq"],
                    "success": True,
                    "command": "runInTerminal",
                    "body": {
                        "shellProcessId": process_id
                    }
                }
                self.io.write_json(response)
                return True
        return False

    def set_breakpoint(self, path : str, lines : list):
        breakpoint_request = {
            "seq": self.new_seq(),
            "type": "request",
            "command": "setBreakpoints",
            "arguments": {
                "source": {
                    "name": path,
                    "path": path
                },
                "lines": lines,
                "breakpoints": [
                    {
                        "line": line
                    } for line in lines
                ],
                "sourceModified": False
            }
        }
        self.io.write_json(breakpoint_request)

    def set_function_breakpoint(self, names : list):
        breakpoint_request = {
            "seq": self.new_seq(),
            "type": "request",
            "command": "setFunctionBreakpoints",
            "arguments": {
                "breakpoints": [
                    {
                        "name": name
                    } for name in names
                ]
            }
        }
        self.io.write_json(breakpoint_request)

    def configuration_done(self):
        complete_request = {
            "seq": self.new_seq(),
            "type": "request",
            "command": "configurationDone"
        }
        self.io.write_json(complete_request)

    def get_stackframes(self, thread_id : int = 1, levels : int = 100) -> list:
        stackframe_request = {
```

```
            "seq": self.new_seq(),
            "type": "request",
            "command": "stackTrace",
            "arguments": {
                "threadId": thread_id,
                "startFrame": 0,
                "levels": levels
            }
        }
        self.io.write_json(stackframe_request)
        output = self.wait("response", command="stackTrace")
        return output["body"]["stackFrames"]

    def next_breakpoint(self, thread_id : int = 1):
        continue_request = {
            "seq": self.new_seq(),
            "type": "request",
            "command": "continue",
            "arguments": {
                "threadId": thread_id
            }
        }
        self.io.write_json(continue_request)

    def step(self, thread_id : int = 1):
        step_request = {
            "seq": self.new_seq(),
            "type": "request",
            "command": "next",
            "arguments": {
                "threadId": thread_id
            }
        }
        self.io.write_json(step_request)

    def step_out(self, thread_id : int = 1):
        step_request = {
            "seq": self.new_seq(),
            "type": "request",
            "command": "stepOut",
            "arguments": {
                "threadId": thread_id
            }
        }
        self.io.write_json(step_request)

    def get_scopes(self, frame_id : int) -> list:
        scopes_request = {
            "seq": self.new_seq(),
            "type": "request",
            "command": "scopes",
            "arguments": {
                "frameId": frame_id
            }
        }
        self.io.write_json(scopes_request)
        output = self.wait("response", command="scopes")
        return output["body"]["scopes"]

    def get_variables(self, scope_id: int) -> list:
        variables_request = {
            "seq": self.new_seq(),
            "type": "request",
            "command": "variables",
            "arguments": {
                "variablesReference": scope_id
            }
        }
```

```python
        }
        self.io.write_json(variables_request)
        output = self.wait("response", command="variables")
        return output["body"]["variables"]

    def evaluate(self, expression : str, frame_id : int, context : str = "repl") -> dict:
        evaluate_request = {
            "seq": self.new_seq(),
            "type": "request",
            "command": "evaluate",
            "arguments": {
                "expression": expression,
                "frameId": frame_id,
                "context": context
            }
        }
        self.io.write_json(evaluate_request)
        return self.wait("response", command="evaluate")

    def wait(self, type: str, event : str = "", command : str = "") -> dict:
        while True:
            output : dict = self.io.read_json() # type: ignore
            if self.debug: print(output, flush=True)
            if output["type"] == "request" and output["command"] == "runInTerminal":
                if self._handleRunInTerminal(output):
                    continue
            if output["type"] == type:
                if event == "" or output["event"] == event:
                    if command == "" or output["command"] == command:
                        return output
            if output["type"] == "event" and output["event"] == "terminated":
                raise DebuggeeTerminatedError()
```

## D    Full Implementation of the Python Probe Server

```python
import os
import subprocess
import sys
import debugpy
from debugpy.common.messaging import JsonIOStream
from livefromdap.utils.StackRecording import Stackframe, StackRecording

from .BaseLiveAgent import BaseLiveAgent

class PythonLiveAgent(BaseLiveAgent):
    def __init__(self, *args, **kwargs):
        super().__init__(*args, **kwargs)
        self.runner_path = kwargs.get("runner_path", os.path.join(os.path.dirname(__file__), "..", "runner", "py_runner.py
            ↪ "))
        self.debugpy_adapter_path = kwargs.get("debugpy_adapter_path", os.path.join(os.path.dirname(debugpy.
            ↪ __file__), "adapter"))

    def start_server(self):
        self.server = subprocess.Popen(
            ["python", self.debugpy_adapter_path],
            stdin=subprocess.PIPE,
            stdout=subprocess.PIPE,
            stderr=subprocess.PIPE,
            restore_signals=False,
            start_new_session=True,
        )
```

```python
        self.io = JsonIOStream.from_process(self.server)

    def restart_server(self):
        self.server.kill()
        self.start_server()

    def stop_server(self):
        self.server.kill()
        if getattr(self, "debugee", None) is not None:
            self.debugee.kill()

    def initialize(self):
        init_request = {
            "seq": self.new_seq(),
            "type": "request",
            "command": "initialize",
            "arguments": {
                "clientID": "vscode",
                "clientName": "Visual Studio Code",
                "adapterID": "python",
                "pathFormat": "path",
                "linesStartAt1": True,
                "columnsStartAt1": True,
                "supportsVariableType": True,
                "supportsVariablePaging": True,
                "supportsRunInTerminalRequest": True,
                "locale": "en",
                "supportsProgressReporting": True,
                "supportsInvalidatedEvent": True,
                "supportsMemoryReferences": True,
                "supportsArgsCanBeInterpretedByShell": True,
                "supportsMemoryEvent": True,
                "supportsStartDebuggingRequest": True
            }
        }
        launch_request = {
            "seq": self.new_seq(),
            "type": "request",
            "command": "launch",
            "arguments": {
                "name": f"Debug Python agent live",
                "type": "python",
                "request": "launch",
                "program": self.runner_path,
                "console": "internalConsole",
                # get the current python interpreter
                "python": sys.executable,
                "debugAdapterPython": sys.executable,
                "debugLauncherPython": sys.executable,
                "clientOS": "unix",
                "cwd": os.getcwd(),
                "envFile": os.path.join(os.getcwd(), ".env"),
                "env": {
                    "PYTHONIOENCODING": "UTF-8",
                    "PYTHONUNBUFFERED": "1"
                },
                "stopOnEntry": False,
                "showReturnValue": True,
                "internalConsoleOptions": "neverOpen",
                "debugOptions": [
                    "ShowReturnValue"
                ],
                "justMyCode": False,
                "workspaceFolder": os.getcwd(),
            }
        }
        self.io.write_json(init_request)
```

```python
        self.io.write_json(launch_request)
        self.wait("event", "initialized")
        self.setup_runner_breakpoint()
        self.wait("event", "stopped")
        return 5

def setup_runner_breakpoint(self):
        self.set_breakpoint(self.runner_path, [16])
        self.configuration_done()

def load_code(self, path: str):
        stacktrace = self.get_stackframes()
        frameId = stacktrace[0]["id"]
        self.evaluate(f"set_import('{os.path.abspath(path)}')", frameId)
        self.next_breakpoint()
        self.wait("event", "stopped")

def execute(self, method, args, max_steps=50):
        self.set_function_breakpoint([method])
        stacktrace = self.get_stackframes()
        frameId = stacktrace[0]["id"]
        self.evaluate(f"set_method('{method}',[{','.join(args)}])", frameId)
        # We need to run the debug agent loop until we are on a breakpoint in the target method
        stackrecording = StackRecording()
        while True:
            stacktrace = self.get_stackframes()
            if stacktrace[0]["name"] == method:
                break
            self.next_breakpoint()
            self.wait("event", "stopped")
        # We are now in the function, we need to get all information, step, and check if we are still in the function
        scope = None
        initial_height = None
        i = 0
        while True:
            stacktrace = self.get_stackframes()
            if initial_height is None:
                initial_height = len(stacktrace)
                height = 0
            else:
                height = len(stacktrace) - initial_height
            if stacktrace[0]["name"] != method:
                break
            # We need to get local variables
            if not scope:
                scope = self.get_scopes(stacktrace[0]["id"])[0]
            variables = self.get_variables(scope["variablesReference"])
            stackframe = Stackframe(stacktrace[0]["line"], stacktrace[0]["column"], height, variables)
            stackrecording.add_stackframe(stackframe)
            i += 1
            if i > max_steps:
                # we need to pop the current frame
                self.restart_server()
                self.initialize()
                return "Interrupted", stackrecording
            self.step()
        # We are now out of the function, we need to get the return value
        scope = self.get_scopes(stacktrace[0]["id"])[0]
        variables = self.get_variables(scope["variablesReference"])
        return_value = None
        for variable in variables:
            if variable["name"] == f'(return) {method}':
                return_value = variable["value"]
        for i in range(2): # Needed to reset the debugger agent loop
            self.next_breakpoint()
            self.wait("event", "stopped")
        return return_value, stackrecording
```

### E   Python Keep-Alive Agent

```python
method = None
method_name = None
method_args = None
import_file = None

def set_import(import_fromp):
    global import_file
    import_file = import_fromp

def set_method(import_methodp, method_argsp):
    global method_name, method_args
    method_name = import_methodp
    method_args = method_argsp

if "__main__" in __name__:
    while True:
        if import_file is not None:
            with open(import_file, "rb") as source_file:
                code = compile(source_file.read(), import_file, "exec")
            exec(code)
            import_file = None
        if method_name is not None and method_args is not None:
            try:
                method = eval(method_name)
                res = method(*method_args)
            except Exception as e:
                pass
            method_args = None
            method_name = None
```

### References

[1]   Andrea Arcuri, Muhammad Zohaib Iqbal, and Lionel Briand. "Random Testing: Theoretical Results and Practical Implications". In: *IEEE Transactions on Software Engineering* 38.2 (2012). DOI: 10.1109/TSE.2011.121.

[2]   Erwan Bousse, Dorian Leroy, Benoit Combemale, Manuel Wimmer, and Benoit Baudry. "Omniscient Debugging for Executable DSLs". In: *Journal of Systems and Software* 137 (2018). DOI: 10.1016/j.jss.2017.11.025.

[3]   Gilad Bracha. "Making Methods Live". 2013. URL: https://gbracha.blogspot.com/2013/04/making-methods-live.html (visited on 2024-02-02).

[4]   Gilad Bracha, Peter von der Ahé, Vassili Bykov, Yaron Kashai, William Maddox, and Eliot Miranda. "Modules as Objects in Newspeak". In: *ECOOP 2010 - Object-Oriented Programming, 24th European Conference*. Volume 6183. LNCS. Springer, 2010. DOI: 10.1007/978-3-642-14107-2_20.

[5]   Ilinca Ciupa, Andreas Leitner, Manuel Oriol, and Bertrand Meyer. "ARTOO: Adaptive Random Testing for Object-Oriented Software". In: *Proceedings of the 30th International Conference on Software Engineering 2008 (ICSE'08)*. May 2008. DOI: 10.1145/1368088.1368099.

[6]   Koen Claessen and John Hughes. "QuickCheck: A Lightweight Tool for Random Testing of Haskell Programs". In: *ACM SIGPLAN Notices* 35.9 (Sept. 2000). DOI: 10.1145/357766.351266.

[7]   Jonathan Edwards. "Example Centric Programming". In: *ACM SIGPLAN Notices* 39.12 (2004). DOI: 10.1145/1052883.1052894.

[8]   Jonathan Edwards, Tomas Petricek, and Tijs van der Storm. "Live & Local Schema Change: Challenge Problems". In: *CoRR* abs/2309.11406 (2023). DOI: 10.48550/arXiv.2309.11406. arXiv: 2309.11406.

[9]   Adele Goldberg. *Smalltalk-80: The Interactive Programming Environment*. Addison-Wesley, 1984. ISBN: 978-0-201-11372-3.

[10]  Christopher Michael Hancock. "Real-time Programming and the Big Ideas of Computational Literacy". PhD thesis. Massachusetts Institute of Technology, 2003. HDL: 1721.1/61549.

[11]  Marijn Haverbeke. "CodeMirror 5". 2024. URL: https://codemirror.net/5/ (visited on 2024-01-07).

[12]  Jevgeni Kabanov. "JRebel Tool Demo". In: *Proceedings of the Fifth Workshop on Bytecode Semantics, Verification, Analysis and Transformation, Bytecode@ETAPS 2010*. Volume 264. Electronic Notes in Theoretical Computer Science 4. Elsevier, 2010. DOI: 10.1016/j.entcs.2011.02.005.

[13]  Sarfraz Khurshid, Corina S. Păsăreanu, and Willem Visser. "Generalized Symbolic Execution for Model Checking and Testing". In: *Tools and Algorithms for the Construction and Analysis of Systems*. Springer Berlin Heidelberg, 2003. DOI: 10.1007/3-540-36577-X_40.

[14]  Bil Lewis. "Debugging Backwards in Time". In: *Proceedings of the Fifth International Workshop on Automated Debugging (AADEBUG 2003), September 2003, Ghent*. Oct. 2003. DOI: 10.48550/arXiv.cs/0310016. arXiv: cs/0310016v1.

[15]  Sean McDirmid. "Usable Live Programming". In: *ACM Symposium on New Ideas in Programming and Reflections on Software, Onward! 2013, part of SPLASH '13, Indianapolis, IN, USA, October 26–31, 2013*. ACM, 2013. DOI: 10.1145/2509578.2509585.

[16]  Microsoft. "Debug Adapter Protocol". 2021. URL: https://microsoft.github.io/debug-adapter-protocol/ (visited on 2024-01-07).

[17]  Microsoft. "VS Code". 2024. URL: https://code.visualstudio.com/ (visited on 2024-01-07).

[18]  Fabio Niephaus, Patrick Rein, Jakob Edding, Jonas Hering, Bastian König, Kolya Opahle, Nico Scordialo, and Robert Hirschfeld. "Example-Based Live Programming for Everyone: Building Language-Agnostic Tools for Live Programming with LSP and GraalVM". In: *Proceedings of the 2020 ACM SIGPLAN International Symposium on New Ideas, New Paradigms, and Reflections on Programming and Software, Onward! 2020*. ACM, 2020. DOI: 10.1145/3426428.3426919.

[19] Robert O'Callahan, Chris Jones, Nathan Froyd, Kyle Huey, Albert Noll, and Nimrod Partush. "Engineering Record and Replay for Deployability". In: *2017 USENIX Annual Technical Conference, USENIX ATC 2017, Santa Clara, CA, USA, July 12–14, 2017*. System: https://rr-project.org/. USENIX Association, 2017, pages 377–389. ISBN: 978-1-931971-38-6.

[20] Oracle. "Java Debug Interface". 2023. URL: https://docs.oracle.com/javase/8/docs/jdk/api/jpda/jdi/index.html (visited on 2024-02-13).

[21] Guillaume Pothier, Éric Tanter, and José M. Piquer. "Scalable Omniscient Debugging". In: *Proceedings of the 22nd Annual ACM SIGPLAN Conference on Object-Oriented Programming, Systems, Languages, and Applications, OOPSLA 2007*. ACM, 2007. DOI: 10.1145/1297027.1297067.

[22] Patrick Rein, Stefan Ramson, Jens Lincke, Robert Hirschfeld, and Tobias Pape. "Exploratory and Live, Programming and Coding. A Literature Study Comparing Perspectives on Liveness". In: *The Art, Science, and Engineering of Programming* 3.1 (2019). DOI: 10.22152/programming-journal.org/2019/3/1. arXiv: 1807.08578v1 [cs.PL].

[23] Riemer van Rozen. "Cascade: A Meta-Language for Change, Cause and Effect". In: *Proceedings of the 16th ACM SIGPLAN International Conference on Software Language Engineering, SLE 2023, Cascais, Portugal, October 23–24, 2023*. ACM, 2023. DOI: 10.1145/3623476.3623515.

[24] Erik Sandewall. "Programming in an Interactive Environment: the LISP Experience". In: *ACM Computing Surveys* 10.1 (Mar. 1978). DOI: 10.1145/356715.356719.

[25] Tijs van der Storm and Felienne Hermans. "Live Literals". In: *Workshop on Live Programming, LIVE'16*. 2016. URL: https://www.cwi.nl/~storm/livelit/livelit.html (visited on 2024-02-02).

[26] Steven L. Tanimoto. "A Perspective on the Evolution of Live Programming". In: *Workshop on Live Programming, LIVE 2013*. IEEE Computer Society, 2013. DOI: 10.1109/LIVE.2013.6617346.

[27] Steven L. Tanimoto. "VIVA: A Visual Language for Image Processing". In: *Journal of Visual Languages & Computing* 1.2 (June 1990). DOI: 10.1016/S1045-926X(05)80012-6.

[28] Warren Teitelman and Larry Masinter. "The Interlisp Programming Environment". In: *IEEE Computer* 14.4 (1981). DOI: 10.1109/C-M.1981.220410.

[29] Ulyana Tikhonova, Jouke Stoel, Tijs van der Storm, and Thomas Degueule. "Constraint-based run-time state migration for live modeling". In: *Proceedings of the 11th ACM SIGPLAN International Conference on Software Language Engineering, SLE 2018, Boston, MA, USA, November 05–06, 2018*. ACM, 2018. DOI: 10.1145/3276604.3276611.

[30] Unknown. "A History of Live Programming". 2013. URL: https://liveprogramming.github.io/liveblog/2013/01/a-history-of-live-programming/ (visited on 2024-02-02).

[31]   Bret Victor. "Inventing on Principle". 2012. URL: https://www.youtube.com/watch?v=PUv66718DII (visited on 2024-02-13).

[32]   Thomas Würthinger. "Graal and Truffle: Modularity and Separation of Concerns as Cornerstones for Building a Multipurpose Runtime". In: *13th International Conference on Modularity, MODULARITY '14*. ACM, 2014. DOI: 10.1145/2584469.2584663.

[33]   Thomas Würthinger, Christian Wimmer, and Lukas Stadler. "Unrestricted and safe dynamic code evolution for Java". In: *Science of Computer Programming* 78.5 (2013). DOI: 10.1016/j.scico.2011.06.005.

## About the authors

**Jean-Baptiste Döderlein** is a Master student at ENS Rennes. He performed this research during an internship at CWI, Amsterdam. Contact him at jean-baptiste.doderlein@ens-rennes.fr.
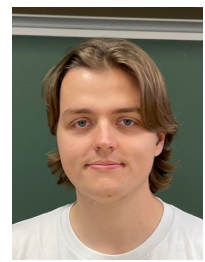
🔵 https://orcid.org/0000-0002-9741-8571

**Riemer van Rozen** is a researcher at CWI who is committed to making programming *"more fun, visual and for everyone"*. He conducts applied research on languages and tools that help create better code more quickly. His research focuses on generic solutions (meta-languages) for domain-specific languages and live programming environments in general, and automated game design in particular. Contact him at rozen@cwi.nl.
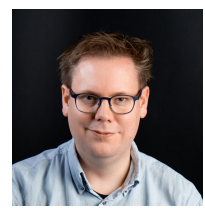
🔵 https://orcid.org/0000-0002-3834-682X

**Tijs van der Storm** is senior researcher at CWI where he heads the Software Analysis & Transformation (SWAT) group, and full professor in software engineering at the University of Groningen, Groningen. His research is centered around the questions: how to make better programming languages? and how to better make programming languages? He is co-designer of the Rascal meta programming system and language workbench. Contact him at storm@cwi.nl.
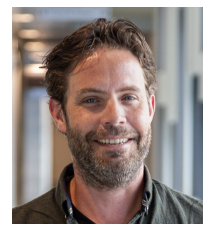
🔵 https://orcid.org/0000-0001-8853-7934